\documentclass{mem}

\usepackage{natbib}\usepackage{txfonts}\usepackage{balance}
\usepackage{graphicx}
\usepackage[a4paper]{hyperref}

\idline{0}{1}

\begin{document}

\title{Granulation across the HR diagram}

\author{
I. Ram\'{\i}rez \inst{1}          \and
C. Allende Prieto \inst{2}        \and
D. L. Lambert \inst{3}               \and
L. Koesterke \inst{3,4}             \and
M. Asplund \inst{1}
}

\offprints{I. Ram\'{\i}rez}

\institute{
Max Planck Institute for Astrophysics, Postfach 1317, 85741 Garching, Germany
\and
Mullard Space Science Laboratory, University College London, UK
\and
McDonald Observatory and Dept. of Astronomy, University of Texas at Austin, USA
\and
Texas Advanced Computing Center, University of Texas, USA \\
\email{ivan@mpa-garching.mpg.de}
}

\authorrunning{Ram\'{\i}rez et al.}

\titlerunning{Granulation across the HR diagram}

\abstract{
We have obtained ultra-high quality spectra ($R=180,000$; $S/N>300$) with unprecedented wavelength coverage (4400 to 7400\,\AA) for a number of stars covering most of the HR diagram in order to test the predictions of models of stellar surface convection. Line bisectors and core wavelength shifts are both measured and modeled, allowing us to validate and/or reveal the limitations of state-of-the-art hydrodynamic model atmospheres of different stellar parameters. We show the status of our project and preliminary results.
}
\maketitle{}

\section{Motivation}

Convection is one of the least understood phenomena in stellar astrophysics. This is, in part, due to the lack of high quality data that can be used to test theoretical models. Such tests can allow us to better understand the physics involved in the convection phenomenon and may help us to reveal the limitations of state-of-the-art models. In addition to this very fundamental motivation, we note that an accurate determination of abundances and fundamental parameters relies on models providing an adequate description of the outer layers of stars. Unrealistic models may lead to uncertain abundances and these errors can easily propagate to our interpretations of Galactic chemical evolution as well as stellar and primordial nucleosynthesis. The impact of granulation on stellar abundances has been shown to be non-negligible in many cases \citep[e.g.,][]{a05}.

\begin{figure*}[t!]
\centering
\includegraphics[width=10cm,bb=50 370 480 640]{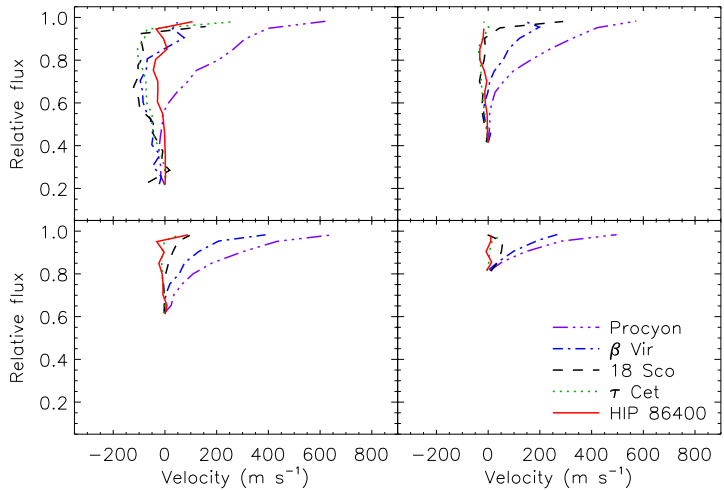}
\caption{\footnotesize Average bisectors of Fe~\textsc{i} lines sorted in four groups of similar line-strength for 4 of our sample dwarf stars and the subgiant star Procyon, covering roughly the temperature range from 4800 to 6600\,K. The cores of these bisectors have been set to be at zero velocity in order to reduce the noise introduced by uncertain laboratory wavelengths \citep[cf.][]{r08,r09}.}
\label{f:bistrend_ms1}
\end{figure*}

\begin{figure*}[t!]
\includegraphics[width=13.7cm,bb=65 400 620 1070]{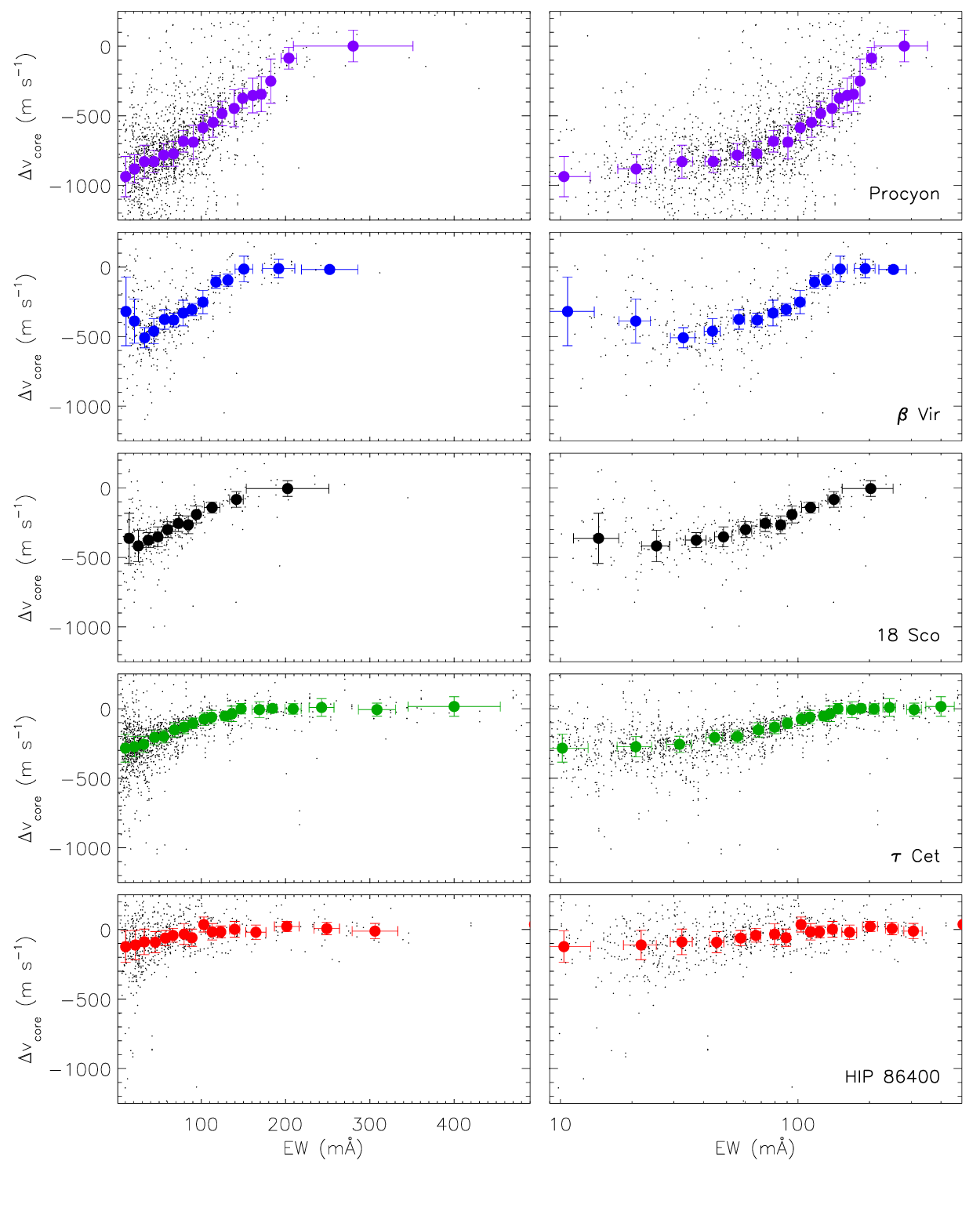}
\caption{\footnotesize Left panels: Core wavelength shifts as a function of equivalent width (EW) of Fe I lines in the spectra of four of our sample dwarf stars and the subgiant star Procyon, covering roughly the temperature range from 4800 to 6600\,K. Right panels: As in the left panels but with a logarithmic ordinate axis scale. Dots correspond to individual lines; filled circles with error bars represent bin-averaged values. The range of convective blueshifts increases from about 200 to 1000 m\,s$^{-1}$ from the K-dwarf HIP\,86400 to the F-subgiant Procyon. The blueshifts decrease for stronger lines but reach a saturation value (a plateau best seen in the left panels, forced to be at zero velocity) that begins at systematically higher EW for warmer stars. Note also that a plateau seems to be present at the low EW end (best seen in the right panels).}
\label{f:lshtrend_ms1}
\end{figure*}

\section{Sample and strategy}

Our sample consists of 15 stars covering a substantial part of the HR diagram and includes one metal-poor giant. Observations are being completed with the Robert~G.~Tull coud\'e spectrograph on the 2.7\,m Harlan~J.~Smith Telescope at McDonald Observatory \citep{t95}. The average spectral resolution of our data is 180,000 and signal-to-noise ratios range from about 300 to 500. The spectra cover (completely in the majority of cases) most of the visible range (4400 to 7400\,\AA) being thus the most complete data set of this quality collected to date.

Line bisectors and core wavelength shifts are measured for unblended Fe~\textsc{i} features listed in the \cite{n94} catalog. Line bisectors of features of similar line-depth are averaged to reduce the noise given that the shape of granulation bisectors is determined mainly by the line strength. These average line bisectors are then compared to their theoretical counterparts computed from 3D hydrodynamic models. If the comparison shows reasonably good agreement, the model is considered valid and can be reliably used in the derivation of 3D corrections to stellar parameters and abundances.

We compute a few lines that have very accurate laboratory data available using both 1D and \textit{validated} 3D models to estimate 3D corrections to chemical abundances derived with the standard 1D approach. We also compute the full spectrum using extensive atomic and molecular line databases to look for features that are potentially sensitive to 3D effects as well as to explore the impact of surface inhomogeneities on absolute fluxes (for example to study their impact on synthetic colors). The necessary tools to compute spectra from 3D hydrodynamic models have already been developed \citep[e.g.,][]{k08}.

As of this writing, we have successfully applied the procedure described above to a sample of K-dwarf stars \citep{r08,r09}. We are currently working on an extension to stars of different stellar parameters. Most of the needed observational material is already at hand. Below, observational results for a few dwarf stars are briefly summarized.

\section{Results: the main-sequence}

Preliminary results for 5 of our sample stars are shown in Figs.~\ref{f:bistrend_ms1} and \ref{f:lshtrend_ms1}. These stars roughly cover the main-sequence from mid-F to early-K spectral type (note however that Procyon is a subgiant).

The trends observed are in good agreement with those found by other authors \citep[e.g.,][]{d87,g05}. The line bisector span increases with effective temperature; those for the strongest lines go from less than 100 m\,s$^{-1}$ in the K-dwarf HIP\,86400 to about 0.6 km\,s$^{-1}$ in Procyon (Fig.~\ref{f:bistrend_ms1}). Weaker lines show smaller bisector span. The bisectors have the typical C-shape that is associated with the granulation phenomenon.

Core wavelength shifts behave similarly. Convective blueshifts are larger (i.e., wavelength shifts are more negative) for the weaker lines of warmer stars (Fig.~\ref{f:lshtrend_ms1}). This is due to the fact that such lines form in deeper atmospheric layers where the granulation contrast is larger and because convection is less efficient in warmer stellar atmospheres, which means that larger velocities are necessary to transport the convective energy.

Along the main-sequence, the effective temperature seems to be the only parameter determining the shapes of line bisectors and the core wavelength shift vs. line strength relations but note that \cite{r08,r09}, who studied these effects in K-dwarfs, have found that stars showing high levels of chromospheric activity presented anomalies. The very active star $\epsilon$ Eri, for example, shows almost vertical line bisectors and a very noisy core wavelength shift vs. EW relation compared to other K-dwarfs even though its spectrum is of the highest quality.

\begin{figure}
\includegraphics[width=7cm,bb=75 370 340 545]{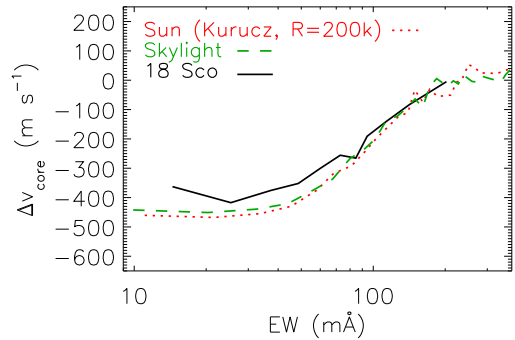}
\caption{\footnotesize Bin-averaged core wavelength shift vs. equivalent width relation (cf. Fig.~\ref{f:lshtrend_ms1}) for the solar spectrum by Kurucz et al. (1984), degraded to a spectral resolution similar to that obtained in our work (red dotted line) as well as our skylight (green dashed line) and 18 Sco (black solid line) observations.}
\label{f:sol_compare2_1}
\end{figure}

In Fig.~\ref{f:sol_compare2_1} we compare the core wavelength shift vs. EW relation obtained using the solar spectrum by \cite{k84} as well as our skylight observations with that corresponding to the solar-twin star 18 Sco \citep{p97}. The latter has fundamental parameters essentially identical to those of the Sun and thus one would expect it to show identical granulation signatures. However, it is clear that the convective blueshifts for the weakest lines in the 18 Sco spectrum are at least 50~m\,s$^{-1}$ too small (less negative) compared to those for the Sun. 

It is interesting to note that the data for 18~Sco were obtained at an epoch when the star was experiencing a maximum of activity (May of 2007), significantly stronger than the Sun's  \citep{h07}. Whether some activity related effect is responsible for this difference needs to be confirmed by follow-up observations of the star during a minimum of chromospheric emission along with nearly simultaneous (and homogeneous; i.e., same spectrograph) observations of a very bright asteroid or other sunlight-reflecting object.

\section{Summary}

We have been able to detect, quantify, and successfully model the granulation signatures present in K-dwarf spectra. Moreover, we have used our validated simulation to explore the impact of 3D effects on stellar abundances and fundamental parameters. Now we plan to extend this work to stars of different stellar parameters covering most of the HR diagram, including a metal-poor giant.

Preliminary results for dwarf stars confirm previous findings that the granulation effects are more important for warmer stars. We find that the effective temperature is possibly not the only parameter determining the strength of these signatures as the comparison of observational results for active and inactive K-type dwarf stars of essentially the same temperature revealed important differences. Furthermore, the granulation signatures of the solar-twin star 18~Sco at its activity peak differ significantly from those observed in the Sun.



\bibliographystyle{aa}

\end{document}